\documentclass[prb,onecolumn,aps,showpacs]{revtex4}
\usepackage{graphicx}%
\usepackage{dcolumn}
\usepackage{amsmath}
\usepackage{amssymb}
\usepackage{epsfig}

\begin{document}

\title{ Strength of the interactions in $YBa_2Cu_3O_{6.7}$ obtained from the
inelastic neutron-scattering measurements}

\author{Z. G. Koinov}\affiliation{Department of Physics and Astronomy,
University of Texas at San Antonio, San Antonio, TX 78249, USA}
\email{Zlatko.Koinov@utsa.edu} 
\begin{abstract}
It is widely accepted that: (i) the angle-resolved photoemission
spectroscopy (ARPES) data produce evidences for the opening of a
d-wave pairing gap in cuprates compounds described at low energies
and temperatures by a BCS theory, and  (ii) the basic pairing
mechanism arises from the antiferromagnetic exchange correlations,
but the charge fluctuations associated with double occupancy of a
site also play an essential role in doped systems. The simplest
model that is consistent with the two statements is the
$t$-$U$-$V$-$J$ model. We have shown that the inelastic neutron
scattering data  on $YBa_2Cu_3O_{6.7}$ [Phys. Rev. Lett.
\textbf{83}, 608 (1999)] combined with the corresponding ARPES data
allow us to obtain the strength of the on-site repulsive interaction
$U$, as well as the strengths of the spin-independent attractive
interaction $V$ and the spin-dependent antiferromagnetic interaction
$J$.
 \end{abstract}

\maketitle

 \section{Introduction} The magnetic susceptibility probed
by neutron scattering in cuprates compounds has provided a variety
of experimental peaks associated with incommensurate and
commensurate structure. This neutron data remains of central
importance in the field of high-Tc superconductivity because the
technique could be used to cover momentum and frequency range which
is wider than that  of any other spectroscopies. The neutron data in
cuprates compounds are characterized by a very strong dependence on
energy, momentum transfer, temperature, and doping. For example, the
existence of a commensurate peak at $\sim 40$ meV and incommensurate
peaks at $\sim 24$ meV and $\sim 32$ meV  has been reported in
underdoped $YBa_2Cu_3O_{6.7}$ \cite{Ar}. The simplest and most
transparent hypothesis put forward by many authors \cite{Ar,Hy} is
that the commensurate resonance and the incommensurate peaks in
cuprates compounds have a common origin. This approach (also known
as the Fermi-liquid approach) is based on the assumption that the
electrons near the Fermi surface are strongly correlated due to the
on-site Hubbard and nearest-neighbor antiferromagnetic interactions.
Alternatively, according to the stripes model (see Ref.
[\onlinecite{Str}] and the references therein) the incommensurate
peaks are the natural descendants of the stripes, which are complex
patterns formed by electrons confined to separate linear regions in
the crystal.

In this paper we analyze the position of commensurate and
incommensurate peaks assuming the Fermi-liquid-based scenario which
consists of two steps. The first step is to obtain the tight-binding
form of mean-field quasiparticle energy and the corresponding
chemical potential by matching the shape of the Fermi surface
measured by the ARPES. Another input parameter for the d-wave
superconductivity is the maximum gap which usually is assumed to be
equal to the ARPES antinodal gap. During the second step of our
approach we use the Bethe-Salpeter (BS) equations to identify
effective interaction strengths $U,V$ and $J$ that are consistent
with both ARPES measurements and the energy and the position of the
commensurate and incommensurate resonances observed by neutron
scattering experiments.

  It is known that
the YBaCuO is a two-layer material, but most of the peak structures
associated with the neutron cross section can be captured by one
layer band calculations \cite{DD}. The effects due to the two-layer
structure can, in principle, be incorporated in our approach, but
this will make the corresponding numerical calculations much more
complicated. In the one-layer approximation, the Hamiltonian of the
$t-J-U-V$ model contains terms representing the hopping of electrons
between sites of the lattice, the on-site repulsive interaction
between two electrons with opposite spins, the attractive
interaction (due to phonons) between electrons on different sites of
the lattice and the spin-dependent Heisenberg near-neighbor
interaction (due to short-range antiferromagnetic order):
\begin{eqnarray}&\widehat{H}=-\sum_{i,j,\sigma}t_{ij}\psi^\dag_{i,\sigma}\psi_{j,\sigma}
-\mu\sum_{i,\sigma}\widehat{n}_{i,\sigma}+U\sum_i
\widehat{n}_{i,\uparrow} \widehat{n}_{i,\downarrow}\nonumber\\&
-V\sum_{<i,j>\sigma\sigma'}\widehat{n}_{i,\sigma}\widehat{n}_{j,\sigma'}+J
\sum_{<i,j>}\overrightarrow{\textbf{S}}_i\textbf{.}\overrightarrow{\textbf{S}}_j,
\label{Hubb1}\end{eqnarray} where $\mu$ is the chemical potential.
The Fermi operator $\psi^\dag_{i,\sigma}$ ($\psi_{i,\sigma}$)
creates (destroys) a fermion on the lattice site $i$ with spin
projection $\sigma=\uparrow,\downarrow$ along a specified direction,
and $\widehat{n}_{i,\sigma}=\psi^\dag_{i,\sigma}\psi_{i,\sigma}$ is
the density operator on site $i$ with a position vector
$\textbf{r}_i$. The symbol $\sum_{<ij>}$ means sum over
nearest-neighbor sites. The first term in (\ref{Hubb1}) is the usual
kinetic energy term in a tight-binding approximation, where $t_{ij}$
is  the single electron hopping integral. We assume $V>0$, so the
fourth term is expected to stabilize the pairing by bringing in a
nearest-neighbor attractive interaction. The last term describes the
nearest-neighbor spin interaction. The spin operator is defined by
$\overrightarrow{\textbf{S}}_i=\psi^\dag_{i,\sigma}
\overrightarrow{\sigma}_{\sigma\sigma'}\psi_{i,\sigma'}/2$, where
$\hbar=1$ and  $\overrightarrow{\sigma}$ is the vector formed by the
Pauli spin matrices $(\sigma_x, \sigma_y,\sigma_z)$. The lattice
spacing is assumed to be $a=1$ and the total number of sites is $N$.

The magnetic neutron scattering directly measures the imaginary part
of the generalized spin susceptibility \cite{L}
$\chi_{ij}(\textbf{Q};\imath
\omega_{p}\rightarrow\omega+\imath\Gamma)$ for momentum transfer
$\textbf{Q}=\textbf{k}_f-\textbf{k}_i$ and energy transfer $\omega$,
where $\textbf{k}_f$ and $\textbf{k}_i$ are the incident and final
neutron wave vectors, respectively.  In high-Tc superconductors,
such as the cuprates compounds, the  phenomenological parameter
$\Gamma$ is a positive function of energy, momentum transfer,
temperature, and doping, and could be measured by the width of the
resonance. Since the position of the peak does not depend on
$\Gamma$ but does depend on the doping, we shall use the available
ARPES data for underdoped $YBa_2Cu_3O_{6.7}$, assuming that
$\Gamma\rightarrow 0^+$. In this case the imaginary part of
$\chi_{ij}$ is a delta function centered at the pole of the real
part of the generalized spin susceptibility. Note that when $\Gamma$
is a positive function, the denominator in the spin susceptibility
vanishes if its real and imaginary parts vanish simultaneously.

The generalized spin susceptibility, related to the Hamiltonian
(\ref{Hubb1}), is defined as follows  \cite{B}:
\begin{equation}\chi_{ij}(\textbf{Q};\imath \omega_{p})=
2\int_0^\beta e^{\imath\omega_p
u}<T_u\{M_i(\textbf{Q},u)M_i(-\textbf{Q},0)\}>du,
\label{As1}\end{equation} where
\begin{eqnarray}&M_i(\textbf{Q},u)=
\frac{1}{2}\left(\sigma_i\right)_{\sigma\sigma'}
\sum_\textbf{q}e^{\widehat{H}u}\psi^\dag_{\textbf{Q}+\textbf{q}/2,\sigma}
\psi_{\textbf{Q}-\textbf{q}/2,\sigma'}e^{-\widehat{H}u}.\nonumber
 \end{eqnarray}
 Here $\psi_{\textbf{q},\sigma}=(1/N)\sum_i e^{\imath\textbf{q.}\textbf{r}_i}\psi_{i,\sigma}$,
 the variable $u$ ranges from $0$ to $\beta=1/k_BT$ ($T$ and
$k_B$  are the temperature and the Boltzmann constant), and
$\omega_{p}=(2\pi/ \beta)p ; p=0, \pm 1, \pm 2,...$. From Eq.
(\ref{As1}) follows that the generalized spin susceptibility and the
two-particle Green's function share common poles.

 The commensurate and incommensurate peak structures
associated with the neutron cross section in YBaCuO have been
studied within the Fermi-liquid-based scenario using the single-band
Hubbard $t-U$ model \cite{Mc,Sch,Er} or the $t-J$ model
\cite{MW,BL,BL1,Li,Li1,YM}. The techniques that have been used are
based on (i) the Monte Carlo numerical calculations \cite{Mc}, (ii)
the random phase approximation (RPA) for the magnetic susceptibility
\cite{Sch,Er}, (iii) the mean-field approximation \cite{MW,BL} and
(iv) the RPA combined with the slave-boson mean field scheme
\cite{BL1,Li,Li1,YM}. In the RPA the two-particle Green's
 function is replaced by a product of two single-particle zero-temperature
  Green's functions \cite{BS} to obtain
\begin{equation}\chi(\textbf{Q};\omega)=\chi^{(0)}(\textbf{Q};
\omega)/(1+U_\textbf{Q}\chi^{(0)}(\textbf{Q};\omega)),\label{RPA}\end{equation}
where the BCS susceptibility $\chi^{(0)}(\textbf{Q};\omega)$  is
given by the usual expression (in our notations
$\chi^{(0)}(\textbf{Q};
\omega)=I_{\widetilde{\gamma}\widetilde{\gamma}}$)
\begin{equation}\chi^{(0)}(\textbf{Q};
\omega)=\frac{1}{2}\sum_{\textbf{k}}\left[1-\frac{\overline{\varepsilon}_\textbf{k}
\overline{\varepsilon}_{\textbf{k}+\textbf{Q}} +\Delta_\textbf{k}
\Delta_{\textbf{k}+\textbf{Q}}}{E(\textbf{k})E(\textbf{k}+\textbf{Q})}\right]
\left[\frac{E(\textbf{k})+E(\textbf{k}+\textbf{Q})}{(\omega+\imath\Gamma)^2
-[E(\textbf{k})+E(\textbf{k}+\textbf{Q})]^2}\right],\label{LF}
\end{equation} and $U_\textbf{Q}=U$ for the Hubbard
model, and $U_\textbf{Q}=-2J(\cos Q_x+\cos Q_y)$ for the $t-J$
model. Here $\overline{\varepsilon}_{\textbf{k}}$ is the mean-field
electron energy , $\Delta_\textbf{k}$ is the gap function and
$E(\textbf{k})=\sqrt{\overline{\varepsilon}^2_\textbf{k}+\Delta^2_\textbf{k}}$.

   Since
there is a consensus that the calculations based upon equation
(\ref{RPA}) overestimate spin fluctuations because the RPA neglects
the mixing between the spin channel and other channels. The coupling
of the spin and two $\pi$ channels (a three-channel
response-function theory) leads in the generalized random phase
approximation (GRPA) to a set of three coupled equations. When the
extended spin channel is added to the previous three channels, we
have a set of four coupled equations (a four-channel theory)
\cite{Plas}. In what follows, the energy of the resonances are
obtained from the solution of 20 coupled Bethe-Salpeter (BS)
equations for the collective modes in GRPA, i.e.  the resonances
emerge due to the mixing between the spin channel and other 19
channels.

\section{Bethe-Salpeter equations for the collective modes}
 The interaction
part of the $t-U-V-J$  Hamiltonian is quartic in the Grassmann
fermion fields so the functional integrals cannot be evaluated
exactly. However, we can transform the quartic terms to a quadratic
form by applying the Hubbard-Stratonovich transformation for the
electron operators \cite{ZK1}:
\begin{eqnarray}&
\int DA
\exp\left[\frac{1}{2}A_{\alpha}(z)D_{\alpha\beta}^{(0)-1}(z,z')A_{\beta}(z)+\widehat{\overline{\psi}}
(y)\widehat{\Gamma}^{(0)}_{\alpha}(y;x|z)\widehat{\psi}(x)A_{\alpha}(z)
\right] =\nonumber\\&\exp\left[-\frac{1}{2}\widehat{\overline{\psi}}
(y)\widehat{\Gamma}^{(0)}_{\alpha}(y;x|z)\widehat{\psi}(x)
D_{\alpha\beta}^{(0)}(z,z') \widehat{\overline{\psi}}
(y')\widehat{\Gamma}^{(0)}_{\beta}(y';x'|z')\widehat{\psi}(x')\right].\nonumber\end{eqnarray}
  The $t-U-V-J$ model has
 an additional term, the spin-dependent interaction Heisenberg
 interaction $J
\sum_{<i,j>}\overrightarrow{\textbf{S}}_i\textbf{.}\overrightarrow{\textbf{S}}_j=J_1+J_2$,
which consists of two terms:
$J_1=\frac{J}{4}\sum_{<i,j>}[\widehat{n}_{i,\uparrow}\widehat{n}_{j,\uparrow}
+\widehat{n}_{i,\downarrow}\widehat{n}_{j,\downarrow}-
\widehat{n}_{i,\uparrow}\widehat{n}_{j,\downarrow}-
\widehat{n}_{i,\downarrow}\widehat{n}_{j,\uparrow}]$ and
$J_2=\frac{J}{2}\sum_{<i,j>}\left[\psi^\dag_{i,\uparrow}\psi_{i,\downarrow}
\psi^\dag_{j,\downarrow}\psi_{j,\uparrow}
+\psi^\dag_{i,\downarrow}\psi_{i,\uparrow}
\psi^\dag_{j,\uparrow}\psi_{j,\downarrow}\right]$. This requires to
introduce four-component Nambu fermion fields
$$\widehat{\overline{\psi}}
(y)=\left(\psi^\dag_\uparrow(y)\psi^\dag_\downarrow(y)\psi_\uparrow(y)\psi_\downarrow(y)
\right)$$  and
$\widehat{\psi}(x)=\left(\psi^\dag_\uparrow(x)\psi^\dag_\downarrow(x)\psi_\uparrow(x)
\psi_\downarrow(x)\right)^T$, where $x$ and $y$ are composite
variables and the field operators obey anticommutation relations.
The  Fourier transforms of the $4\times 4$ matrices
$\widehat{D}_{\alpha\beta}^{(0)}$ and
$\widehat{\Gamma}^{(0)}_{\alpha}$ ($\alpha,\beta=1,2,3,4$) can be
written in terms of the Pauli $\sigma_i$, Dirac $\gamma^0$ and alpha
matrices \cite{B}:
$\widehat{D}^{(0)}=\left(\begin{array}{cc}\widehat{D}_1&0\\0&\widehat{D}_2\end{array}%
\right)$, $\widehat{\Gamma}_{1,2}^{(0)}=(\gamma^0\pm\alpha_z)/2$ and
 $\widehat{\Gamma}_{3,4}^{(0)}=(\alpha_x\pm \imath\alpha_y)/2$, where
$\alpha_i=\left(\begin{array}{cc}\sigma_i&0\\0&\sigma_y\sigma_i\sigma_y
\end{array}%
\right)$,  $\widehat{D}_1=
\left(J(\textbf{k})-V(\textbf{k})\right)\sigma_0+\left(U-J(\textbf{k})-V(\textbf{k})\right)
\sigma_x$ and $\widehat{D}_2=2J(\textbf{k})\sigma_x$.   For a square
lattice and nearest-neighbor interactions $V(\textbf{k})=4V(\cos
k_x+\cos k_y)$ and $J(\textbf{k})=J(\cos k_x+\cos k_y)$.  Now, we
can establish a one-to-one correspondence between the system under
consideration and a system
 which consists of a four-component boson field
$A_{\alpha}(z)$ interacting with fermion fields
$\widehat{\overline{\psi}} (y)$ and $\widehat{\psi}(x)$. The action
of the model system is $S= S^{(e)}_0+S^{(A)}_0+S^{(e-A)}$ where:
$S^{(e)}_0=\widehat{\overline{\psi}
}(y)\widehat{G}^{(0)-1}(y;x)\widehat{\psi} (x)$, $
S^{(A)}_0=\frac{1}{2}A_{\alpha}(z)D^{(0)-1}_{\alpha
\beta}(z,z')A_{\beta}(z')$ and $ S^{(e-A)}=\widehat{\overline{\psi}}
(y)\widehat{\Gamma}^{(0)}_{\alpha}(y,x\mid z)\widehat{\psi}
(x)A_{\alpha}(z)$. Here, we have used composite variables
$x,y,z=\{\textbf{r}_i,u\}$, where $\textbf{r}_{i}$ is a lattice site
vector, and  variable $u$ range from $0$ to $\beta=1/k_BT$ ($T$ and
$k_B$  are the temperature and the Boltzmann constant). We set
$\hbar=1$ and we use the summation-integration convention: that
repeated variables are summed up or integrated over.

 In Ref. [\onlinecite{ZK1}], the spectrum of the collective excitations of the extended Hubbard model has
 been obtained from the Dyson equation for the boson Green's
 function $D$ in terms of the proper self-energy. In what follows we
 shall obtain the spectrum of the collective excitations directly from
 the solutions of the BS equations for the two-particle Green's function.
It can be shown that for a singlet superconductivity and d-wave
pairing $J_1$ and $J_2$ terms contribute separately to the
collective modes. Since the RPA expression (\ref{RPA}) for the spin
response function can be obtained by keeping only the $J_1$
interaction term in our BS equations (\ref{NewEq1}) and
(\ref{NewEq2}) when U=V=0, we shall neglect the contributions to the
BS equations due to the $J_2$ term. Following the same steps as in
Refs. [\onlinecite{CG,ZK}], we can derive  a set of two BS equations
for the collective mode $\omega(\textbf{Q})$ and corresponding BS
amplitudes:
\begin{eqnarray}&[\omega(\textbf{Q})-\varepsilon(\textbf{k},\textbf{Q})]
G^{+}(\textbf{k},\textbf{Q})= \frac{U}{2N}\sum_{\textbf{q}}
\left[\gamma_{\textbf{k},\textbf{Q}}\gamma_{\textbf{q},\textbf{Q}}+
l_{\textbf{k},\textbf{Q}}l_{\textbf{q},\textbf{Q}}\right]
G^{+}(\textbf{q},\textbf{Q})\nonumber\\&-\frac{U}{2N}\sum_{\textbf{q}}
\left[\gamma_{\textbf{k},\textbf{Q}}\gamma_{\textbf{q},\textbf{Q}}-
l_{\textbf{k},\textbf{Q}}l_{\textbf{q},\textbf{Q}}\right]
G^{-}(\textbf{q},\textbf{Q})\nonumber\\&-\frac{1}{2N}\sum_{\textbf{q}}\left[V(\textbf{k}-\textbf{q})+
J(\textbf{k}-\textbf{q})\right]
\left[\gamma_{\textbf{k},\textbf{Q}}\gamma_{\textbf{q},\textbf{Q}}+
l_{\textbf{k},\textbf{Q}}l_{\textbf{q},\textbf{Q}}\right]
G^{+}(\textbf{q},\textbf{Q})\nonumber\\&-\frac{1}{2N}\sum_{\textbf{q}}\left[V(\textbf{k}-\textbf{q})-
J(\textbf{k}-\textbf{q})\right] \left[
\widetilde{\gamma}_{\textbf{k},\textbf{Q}}
\widetilde{\gamma}_{\textbf{q},\textbf{Q}}+m_{\textbf{k},\textbf{Q}}m_{\textbf{q},\textbf{Q}}\right]
G^{+}(\textbf{q},\textbf{Q})\nonumber\\&+\frac{1}{2N}\sum_{\textbf{q}}\left[V(\textbf{k}-\textbf{q})+
J(\textbf{k}-\textbf{q})\right]
\left[\gamma_{\textbf{k},\textbf{Q}}\gamma_{\textbf{q},\textbf{Q}}-
l_{\textbf{k},\textbf{Q}}l_{\textbf{q},\textbf{Q}}\right]
G^{-}(\textbf{q},\textbf{Q})\nonumber\\&+\frac{1}{2N}\sum_{\textbf{q}}\left[V(\textbf{k}-\textbf{q})-
J(\textbf{k}-\textbf{q})\right] \left[
\widetilde{\gamma}_{\textbf{k},\textbf{Q}}
\widetilde{\gamma}_{\textbf{q},\textbf{Q}}-m_{\textbf{k},\textbf{Q}}m_{\textbf{q},\textbf{Q}}\right]
G^{-}(\textbf{q},\textbf{Q})\nonumber\\&-\frac{U-2J(\textbf{Q})}{2N}\sum_{\textbf{q}}
\widetilde{\gamma}_{\textbf{k},\textbf{Q}}\widetilde{\gamma}_{\textbf{q},\textbf{Q}}
\left(G^{+}(\textbf{q},\textbf{Q})
-G^{-}(\textbf{q},\textbf{Q})\right)\nonumber\\&+\frac{U-2V(\textbf{Q})}{2N}\sum_{\textbf{q}}m_{\textbf{k},\textbf{Q}}m_{\textbf{q},\textbf{Q}}
\left[G^{+}(\textbf{q},\textbf{Q})+G^{-}(\textbf{q},\textbf{Q})\right],
\label{NewEq1}
\end{eqnarray}
\begin{eqnarray}
&[\omega(\textbf{Q})+\varepsilon(\textbf{k},\textbf{Q})]G^{-}(\textbf{k},\textbf{Q})=
-\frac{U}{2N}\sum_{\textbf{q}}
\left[\gamma_{\textbf{k},\textbf{Q}}\gamma_{\textbf{q},\textbf{Q}}+
l_{\textbf{k},\textbf{Q}}l_{\textbf{q},\textbf{Q}}\right]
G^{-}(\textbf{q},\textbf{Q})\nonumber\\&+\frac{U}{2N}\sum_{\textbf{q}}
\left[\gamma_{\textbf{k},\textbf{Q}}\gamma_{\textbf{q},\textbf{Q}}-
l_{\textbf{k},\textbf{Q}}l_{\textbf{q},\textbf{Q}}\right]
G^{+}(\textbf{q},\textbf{Q})\nonumber\\&+\frac{1}{2N}\sum_{\textbf{q}}\left[V(\textbf{k}-\textbf{q})+
J(\textbf{k}-\textbf{q})\right]
\left[\gamma_{\textbf{k},\textbf{Q}}\gamma_{\textbf{q},\textbf{Q}}+
l_{\textbf{k},\textbf{Q}}l_{\textbf{q},\textbf{Q}}\right]
G^{-}(\textbf{q},\textbf{Q})\nonumber\\&
+\frac{1}{2N}\sum_{\textbf{q}}\left[V(\textbf{k}-\textbf{q})-
J(\textbf{k}-\textbf{q})\right] \left[
\widetilde{\gamma}_{\textbf{k},\textbf{Q}}
\widetilde{\gamma}_{\textbf{q},\textbf{Q}}
+m_{\textbf{k},\textbf{Q}}m_{\textbf{q},\textbf{Q}}\right]
G^{-}(\textbf{q},\textbf{Q})\nonumber\\&-\frac{1}{2N}
\sum_{\textbf{q}}\left[V(\textbf{k}-\textbf{q})+
J(\textbf{k}-\textbf{q})\right]
\left[\gamma_{\textbf{k},\textbf{Q}}\gamma_{\textbf{q},\textbf{Q}}-
l_{\textbf{k},\textbf{Q}}l_{\textbf{q},\textbf{Q}}\right]
G^{+}(\textbf{q},\textbf{Q})\nonumber\\&-\frac{1}{2N}
\sum_{\textbf{q}}\left[V(\textbf{k}-\textbf{q})-
J(\textbf{k}-\textbf{q})\right] \left[
\widetilde{\gamma}_{\textbf{k},\textbf{Q}}
\widetilde{\gamma}_{\textbf{q},\textbf{Q}}-m_{\textbf{k},\textbf{Q}}
m_{\textbf{q},\textbf{Q}}\right]
G^{+}(\textbf{q},\textbf{Q})\nonumber\\&
-\frac{U-2J(\textbf{Q})}{2N}\sum_{\textbf{q}}
\widetilde{\gamma}_{\textbf{k},\textbf{Q}}\widetilde{\gamma}_{\textbf{q},\textbf{Q}}
\left(G^{+}(\textbf{q},\textbf{Q})
-G^{-}(\textbf{q},\textbf{Q})\right)\nonumber\\&
-\frac{U-2V(\textbf{Q})}{2N}\sum_{\textbf{q}}m_{\textbf{k},\textbf{Q}}m_{\textbf{q},\textbf{Q}}
\left[G^{+}(\textbf{q},\textbf{Q})+G^{-}(\textbf{q},\textbf{Q})\right].
\label{NewEq2}
\end{eqnarray}
Here $\varepsilon(\textbf{k},\textbf{Q})=
E(\textbf{k}+\textbf{Q})+E(\textbf{k})$, and we use the same form
factors as in Ref.[\onlinecite{CG}]:
$\gamma_{\textbf{k},\textbf{Q}}=u_{\textbf{k}}u_{\textbf{k}+\textbf{Q}}+v_{\textbf{k}}v_{\textbf{k}+\textbf{Q}},\quad
l_{\textbf{k},\textbf{Q}}=u_{\textbf{k}}u_{\textbf{k}+\textbf{Q}}-v_{\textbf{k}}v_{\textbf{k}+\textbf{Q}},
\quad
\widetilde{\gamma}_{\textbf{k},\textbf{Q}}=u_{\textbf{k}}v_{\textbf{k}+\textbf{Q}}-u_{\textbf{k}+\textbf{Q}}v_{\textbf{k}},
$ and $ m_{\textbf{k},\textbf{Q}}=
u_{\textbf{k}}v_{\textbf{k}+\textbf{Q}}+u_{\textbf{k}+\textbf{Q}}v_{\textbf{k}}
$ where $u^2_{\textbf{k}}=1-v^2_{\textbf{k}}=
\left[1+\overline{\varepsilon}(\textbf{k})/E(\textbf{k})\right]/2$.

 The Fourier
transforms of  $V$ and $J$ interactions are separable, i.e.
$V(\textbf{k}-\textbf{q})=2V\widehat{\lambda}_\textbf{k}\widehat{\lambda}^T_\textbf{q}$
and
$J(\textbf{k}-\textbf{q})=J\widehat{\lambda}_\textbf{k}\widehat{\lambda}^T_\textbf{q}/2$,
and therefore, Eqs. (\ref{NewEq1}) and (\ref{NewEq2}) can be solved
analytically. Here
$\widehat{\lambda}_\textbf{k}=\left(s_\textbf{k},d_\textbf{k},ss_\textbf{k},sd_\textbf{k}\right)$
 is an $1\times 4$ matrix, and we have used the following notations:
$s_\textbf{k}=\cos(k_x)+\cos(k_y)$,
$d_\textbf{k}=\cos(k_x)-\cos(k_y)$,
$ss_\textbf{k}=\sin(k_x)+\sin(k_y)$ and
$cd_\textbf{k}=\sin(k_x)-\sin(k_y)$. Thus, the BS equations for the
collective modes can be reduced to a set of 20 coupled linear
homogeneous equations. The existence of a non-trivial solution
requires that the secular determinant
$det\|\widehat{\chi}^{-1}-\widehat{V}\|$ is equal to zero, where the
bare mean-field-quasiparticle response function
$\widehat{\chi}=\left(
\begin{array}{cc}
P&Q\\
Q^T&R
\end{array}%
\right)$  and the interaction
$\widehat{V}=diag(U,U,-(U-2J(\textbf{Q})),U-2V(\textbf{Q}),-(2V+J/2),...,-(2V+J/2),-(2V-J/2),...,-(2V-J/2))$
are $20\times 20$ matrices.  Here, $P$ and $Q$ are $4\times 4$ and
$4\times 16$ blocks, respectively, while $R$ is $16\times 16$ block
(in what follows $i,j=1,2,3,4$):\begin{eqnarray}& P=\left|
\begin{array}{cccc}
I_{\gamma,\gamma}&J_{\gamma,l}&I_{\gamma,\widetilde{\gamma}}&J_{\gamma,m}\\
J_{\gamma,l}&I_{l,l}&J_{l,\widetilde{\gamma}}&I_{l,m}\\
I_{\gamma,\widetilde{\gamma}}&J_{l,\widetilde{\gamma}}&
I_{\widetilde{\gamma},\widetilde{\gamma}}&
J_{\widetilde{\gamma},m}\\
J_{\gamma,m}&I_{l,m}&J_{\widetilde{\gamma},m}&I_{m,m}
\end{array}%
\right|,Q=\left|
\begin{array}{cccc}
I^i_{\gamma,\gamma}&J^i_{\gamma,l}&I^i_{\gamma,\widetilde{\gamma}}&J^i_{\gamma,m}\\
J^i_{\gamma,l} &I^i_{l,l}&J^i_{l,\widetilde{\gamma}}&I^i_{l,m}\\
I^i_{\gamma,\widetilde{\gamma}}& J^i_{l,\widetilde{\gamma}}&I^i_{\widetilde{\gamma},\widetilde{\gamma}}&J^i_{\widetilde{\gamma},m}\\
J^i_{\gamma,m}&I^i_{l,m} &J^i_{\widetilde{\gamma},m} &I^i_{m,m}
\end{array}%
\right|,\nonumber\\& R=\left|
\begin{array}{cccc}
I^{ij}_{\gamma, \gamma}&J^{ij}_{\gamma, l}& I^{ij}_{\gamma,
\widetilde{\gamma}}&J^{ij}_{\gamma,
 m}\\
J^{ij}_{\gamma, l} &I^{ij}_{l,l}&
J^{ij}_{l,\widetilde{\gamma}}&I^{ij}_{l,m}\\
 I^{ij}_{\gamma
\widetilde{\gamma}}& J^{ij}_{l,\widetilde{\gamma}}&
 I^{ij}_{\widetilde{\gamma},\widetilde{\gamma}}&
J^{ij}_{\widetilde{\gamma}, m}\\
 J^{ij}_{\gamma,
 m}&I^{ij}_{l,m} &J^{ij}_{\widetilde{\gamma}, m}
  &I^{ij}_{m,m}
\end{array}%
\right|.\nonumber\end{eqnarray}  The quantities
$I_{a,b}=F_{a,b}(\varepsilon (\mathbf{k},\mathbf{Q}))$ and
$J_{a,b}=F_{a,b}(\omega)$, the $1\times 4$ matrices
$I^i_{a,b}=F^{i}_{a,b}(\varepsilon (\mathbf{k},\mathbf{Q}))$ and
$J^i_{a,b}=F^{i}_{a,b}(\omega)$, and the $4\times 4$ matrices
$I^{ij}_{a,b}=F^{ij}_{a,b}(\varepsilon (\mathbf{k},\mathbf{Q}))$ and
$J^{ij}_{a,b}=F^{ij}_{a,b}(\omega)$ are defined as follows (the quantities $a(\mathbf{k}%
,\mathbf{Q})$ and $b(\mathbf{k},\mathbf{Q})=l_{\mathbf{k},\mathbf{Q}},m_{%
\mathbf{k},\mathbf{Q}},\gamma _{\mathbf{k},\mathbf{Q}}$ or $\widetilde{%
\gamma }_{\mathbf{k},\mathbf{Q}}$):
$$
F_{a,b}(x)\equiv \frac{1}{N}\sum_\textbf{k}\frac{%
xa(\mathbf{k},\mathbf{Q})b(\mathbf{k},%
\mathbf{Q})}{\omega ^{2}-\varepsilon ^{2}(\mathbf{k},\mathbf{Q})},
F^i_{a,b}(x)\equiv \frac{1}{N}\sum_\textbf{k}\frac{%
xa(\mathbf{k},\mathbf{Q})b(\mathbf{k},%
\mathbf{Q})\widehat{\lambda}^i_\textbf{k}}{\omega ^{2}-\varepsilon
^{2}(\mathbf{k},\mathbf{Q})},$$ $$F^{ij}_{a,b}(x)\equiv \frac{1}{N}\sum_\textbf{k}\frac{%
xa(\mathbf{k},\mathbf{Q})b(\mathbf{k},%
\mathbf{Q})}{\omega ^{2}-\varepsilon
^{2}(\mathbf{k},\mathbf{Q})}\left(\widehat{\lambda}^T_\textbf{k}
\widehat{\lambda}_\textbf{k}\right)_{ij}.$$

\section{Numerical calculations}
 The sum over
$\textbf{k}$ can be replaced by a double integral over $k_x$ and
$k_y$, both from the first Brillouin zone. After that, we applied
the substitutions $x=\tan k_x/4$ and $y=\tan k_y/4$ to rewrite the
integrals in the form of Gaussian quadrature
$\int_{-1}^{1}dx\int_{-1}^{1}dy f(x,y)/(1+x)(1+y)$. The
corresponding integrals are numerically evaluated using $49\times
49$ $(x_i,y_j)$ points: $\int_{-1}^{1}dx\int_{-1}^{1}dy
f(x,y)/(1+x)(1+y)=\sum_{i=1}^{49}\sum_{j=1}^{49}w_iw_jf(x_i,y_j)$,
where $w_i$ is the corresponding weight.

The mean-field electron energy $\overline{\varepsilon}_\textbf{k}$
has a tight-binding form
\begin{equation}\overline{\varepsilon}_\textbf{k}=-2t\left(\cos k_x+\cos
k_y\right)+4t' \cos k_x\cos k_y -2t''(\cos 2k_x+\cos
2k_y)-\mu\label{En}\end{equation} obtained by fitting the ARPES data
with a chemical potential $\mu$ and hopping amplitudes $t_i$ for
first to third nearest neighbors on a square lattice. Using the
established approximate parabolic relationship \cite{Tr} $T_c
/T_{c,max}=1-82.6(p-0.16)^2$, where $Tc,max\sim 93$K is the maximum
transition temperature of the system, $T_c=67$K is the transition
temperature for underdoped $YBa_2Cu_3O_{6.7}$, we find that  the
hole doping is $p=0.10$. At that level of doping the ARPES
parameters are obtained in Ref. \cite{ARPES}: $t=0.25$ eV,
$t'=0.4t$, $t''=0.0444t$ and $\mu=-0.27$ eV.  In the case of
d-pairing the gap function is $\Delta_\textbf{k}=\Delta
d_\textbf{k}/2$, where the  gap maximum $\Delta$ should agree with
ARPES experiments. In the case of underdoped $YBa_2Cu_3O_{6.7}$ the
gap maximum has to be between the corresponding $\Delta=66$ meV in
$YBa_2Cu_3O_{6.6}$ and $\Delta=50$ meV in $YBa_2Cu_3O_{6.95}$
\cite{Delta}, so we set $\Delta=60$ meV. The BCS gap equation is
$$1=\frac{V_\psi}{2}\int^\pi_{-\pi}\int^\pi_{-\pi}
\frac{d\textbf{k}}{(2\pi)^2} \frac{d^2_\textbf{k}}
{E(\textbf{k})},$$ where $V_\psi=2V+3J/2$. The numerical solution of
the gap equation provides $V_\psi=265$ meV.

\begin{figure}[tbp]\includegraphics{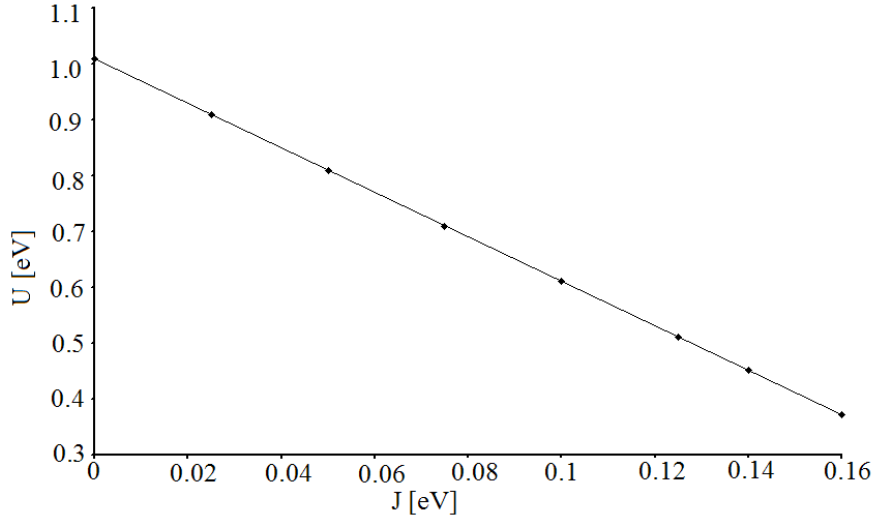} \label{Fig. 1}
\caption{Set of points in $U,J$ parameter space which reproduces the
commensurate resonance at 40 meV at point
$\textbf{Q}_0=(\pi/a,\pi/a$) . The set is fitted with the linear
formula $U=-3.985 J+1.01$ ($U$ and $J$ are in eV). Note that
$V=V_\psi/2-3J/4$ where $V_\psi=265$ meV is calculated from the gap
equation by using the set of parameters given in Ref.
[\onlinecite{ARPES}]. The maximum value of the energy gap is
$\Delta=60$ meV Ref. [\onlinecite{Delta}].}
\end{figure}
During the second step of our approach we solved numerically the BS
equations to obtain the spectrum of the collective modes
$\omega(\textbf{Q})$ at the commensurate point
$\textbf{Q}_0=(\pi/a,\pi/a)$, as well as at four incommensurate
points $\pi/a(1\pm \delta,1)$ and $\pi/a(1,1\pm \delta)$.  Here
$\delta$ is the deviation of the collective mode position from
$\textbf{Q}_0$. The collective-mode energy is the same at any of the
four incommensurate points. Since the spin response function and the
two-particle Green's function share the same poles, we can use the
40 meV solution of the BS equations at $\textbf{Q}_0$ to obtain a
relation between $U$ and $J$ parameters, which is presented in FIG.
1. It can be seen that the linear formula $U=-3.985 J+1.01$ agrees
very well with the numerical results ($U$ and $J$ are in eV). By
means of the last relation we have solved the BS equations for the
one of the four incommensurate 24 meV peaks observed at
$(Q_x,Q_y)=\pi/a(1\pm\delta,1)$ and $(Q_x,Q_y)=\pi/a(1,1\pm\delta)$,
where $\delta=0.22$ \cite{Ar}. The solution provides the following
interaction strengths: $J\sim 129$ meV, $V\sim 35.7$ meV and $U\sim
495$ meV.

We have tested the above values of the interaction strengths by
calculating the positions of the incommensurate peaks at 32 meV. The
Bethe-Salpeter equations with the above strengths provide the
deviation from $\textbf{Q}_0$ of about $\delta=0.196$. This result
is in agreement with the experimentally obtained deviation of
$\delta\sim 0.19$ (see FIG. 2 in Ref. [\onlinecite{Ar}]). The fact
that our approach is able to reproduce the 40 meV peak as well as
the two peaks at 24 meV and 32 meV allows us to conclude that the
commensurate resonance and incommensurate peaks have a common
origin.

\section{Summary and discution}
The strength for $J$ should be comparable to the strength of the
superexchange interactions in the underdoped antiferromagnetic
insulator state of the cuprates. The superexchange interaction in
 cuprates has been studied by using several
experimental tools, and is now known to be not strongly dependent on
materials with the magnitude of $0.1-0.12$ eV. Our value of
$J=0.129$ eV is larger than the accepted experimental values though
some theoretical papers have predicted different magnitudes. For
example, in Ref. [\onlinecite{FJ}] the value of $J=0.22$ eV has been
predicted using standard cuprate parameters. The calculated value of
the superexchange interaction in Ref. [\onlinecite{MT}] is about
$J=0.16$ eV. The differences could be due to the fact that
calculations are very sensitive to the values of the superconducting
gap and the tight-binding parameters, and therefore, somewhat
different tight-binding form of the mean-field electron energy could
bring the calculated $J$ closer to the experimental value.

In summary, we have demonstrated that the strengths of the
interactions in cuprates can be obtained if we have the
angle-resolved photoemission and inelastic neutron scattering data
collected on the same crystals of the high-temperature
superconductor. We do not wish to repeat the theoretical arguments
that were advanced against the stripes model, but our unified
description of the peaks based on the $t-U-V-J$ model strongly
supports the hypothesis that the commensurate resonance and the
incommensurate peaks in cuprates compounds have a common origin.

\end{document}